\documentclass[iop]{emulateapj}

\usepackage{braket}

\shorttitle{Planet Traps and the Critical Metallicity for First Giant Planets}
\shortauthors{Hasegawa \& Hirashita}

\begin{document}

\title{Planet Traps and First Planets: the Critical Metallicity for Gas Giant Formation}

\author{Yasuhiro Hasegawa\altaffilmark{1} and Hiroyuki Hirashita}
\affil{Institute of Astronomy and Astrophysics, Academia Sinica (ASIAA), P.O. Box 23-141, Taipei 10617, Taiwan}
\email{YH: yasu@asiaa.sinica.edu.tw, HH: hirashita@asiaa.sinica.edu.tw}

\altaffiltext{1}{EACOA fellow}

\begin{abstract}
The ubiquity of planets poses an interesting question: when first planets are formed in galaxies. 
We investigate this problem by adopting a theoretical model developed for understanding the statistical properties of exoplanets. 
Our model is constructed as the combination of planet traps with the standard core accretion scenario 
in which the efficiency of forming planetary cores directly relates to the dust density in disks or the metallicity ([Fe/H]).
We statistically compute planet formation frequencies (PFFs) as well as 
the orbital radius ($\braket{R_{rapid}}$) within which gas accretion becomes efficient enough to form Jovian planets. 
The three characteristic planetary populations that are inferred from the accumulation of observed exoplanets are examined: 
hot Jupiters ($0.01 < r_p / \mbox{AU}< 0.5$, $30 < M_p/M_{\oplus} < 10^4$, where $r_p$ and $M_p$ are the semimajor axis and the mass of planets, respectively), 
exo-Jupiters ($0.5 < r_p / \mbox{AU}< 10$, $30 < M_p/M_{\oplus} < 10^4$), 
and low-mass planets such as super-Earths ($0.01 < r_p / \mbox{AU}< 0.5$, $1 < M_p/M_{\oplus} < 30$).
We explore the behavior of the PFFs as well as $\braket{R_{rapid}}$ for the three different populations as a function of metallicity ($-2 \leq$[Fe/H]$\leq -0.6$).
We show that the total PFFs (the sum of the PFFs for all the three populations) increase steadily with metallicity,  
which is the direct outcome of the core accretion picture. 
For the entire range of the metallicity considered here, the population of the low-mass planets dominates over the Jovian planets 
(i.e. the hot and the exo-Jupiters). 
The Jovian planets contribute to the PFFs above [Fe/H]$\simeq -1$. 
For the formation of two kinds of the Jovian planets, we find that the hot Jupiters form at lower metallcities than the exo-Jupiters. 
This arises from the radially inward transport of planetary cores by their host traps. 
Our results show that the transport becomes more effective for disks with lower metallicities due to the slower growth of the cores.
The PFFs for the exo-Jupiters exceed those for the hot Jupiters around [Fe/H]$\simeq -0.7$.
Finally, we show that the critical metallicity for forming Jovian planets is [Fe/H]$\simeq -1.2$,  
which is evaluated by comparing the values of $\braket{R_{rapid}}$ between the hot Jupiters and the low-mass planets.   
The comparison intrinsically links to the different gas accretion efficiency between these two types of planets. 
This study therefore implies that important physical processes in planet formation may be tested by examining exoplanets around metal-poor stars.
\end{abstract}


\section{Introduction} \label{intro}

The unprecedented success in exoplanet observations has inferred that planet formation occurs in various environments \citep[e.g.,][]{us07,mml11,brb12}. 
The metallicity of host stars provides one of the major factors for discovering exoplanets \citep[e.g.,][]{sim04,sl11,blj12}. 
The observations suggest that massive planets like our Jupiters are observed preferentially around metal-rich stars 
whereas the detectability of low-mass planets such as super-Earths apparently does not correlate with stellar metallicity. 
The metallicity dependence is often referred to as the "planet-metallicity relation".
Currently, the presence of exoplanets is confirmed around stars with a wide range of metallicities ([Fe/H]$\la 0.6$).\footnote{
The standard notation is adopted, so that [Fe/H] is a logarithmic metallicity normalized by the solar metallicity. In other words, [Fe/H]=0 for solar metallicity.} 
The lowest metallicity at which massive exoplanets are observed is so far [Fe/H]$\simeq -2$ \citep{skh10,srd12}, 
although these observations are currently a matter of debate. 
In fact, other observations show that the presence of planets for the same targets is very likely to be ruled out \citep{dsb13,mrh13,jj14}.
Most observed massive and low-mass planets are confined well within [Fe/H]$\ga -0.6$.

A successful theoretical framework to understand the exoplanet observations is the so-called core accretion scenario \citep[e.g.,][]{p96,il04i}. 
In this scenario, gas giants undergo sequential accretion of dust and gas: cores of gas giants form first via oligarchic growth by planetesimal collisions 
\citep[e.g.,][]{ki98,tdl03}, and then the cores accrete surrounding gas and form massive envelopes around them \citep[e.g.,][]{m80,ine00,lhdb09,mbp10}.  
Population synthesis calculations developed by this picture confirmed 
that massive planets within the orbital radius $r=10$ AU can be built within the disk lifetime \citep[$\sim 10^{6-7}$ yr, e.g.,][]{il04i,mab09,hp13a}. 
Furthermore, the calculations succeeded well in reproducing the planet-metallicity relation.  
This arises because the efficiency of forming planetary cores is directly linked to the number density of dust in disks in the model 
\citep[e.g.,][]{il04ii,mab12,hp14a}. 
One of the intriguing questions that remain to be addressed is 
whether or not we can extrapolate these results to metal-poor stars ([Fe/H]$< -0.6$).
What is the critical metallicity for forming gas giants in the standard core accretion picture? 

In this paper, we explore the problem and quantify at what value of metallicity the formation of gas giants can proceed. 
To this end, we carry out a statistical analysis for planetary populations by using 
a semi-analytical model developed in a series of earlier papers \citep[hereafter, HP11, HP12, HP13, HP14]{hp11,hp12,hp13a,hp14a}. 
By computing planet formation and migration in evolving gaseous disks, 
we estimate the planet formation frequency (PFF) as well as 
the averaged value of $R_{rapid}$ (defined as $\braket{R_{rapid}}$) within which gas accretion onto the cores becomes rapid ($\sim 10^5$ yr).
The analysis is applied for three different planetary populations that are suggested by the observations (see Table \ref{table1}): 
hot Jupiters ($0.01 < r_p / \mbox{AU}< 0.5$, $30 < M_p/M_{\oplus} < 10^4$, where $r_p$ and $M_p$ are the semimajor axis and the mass of planets, respectively), 
exo-Jupiters ($0.5 < r_p / \mbox{AU}< 10$, $30 < M_p/M_{\oplus} < 10^4$), 
and low-mass planets with short orbital periods such as super-Earths ($0.01 < r_p / \mbox{AU}< 0.5$, $1 < M_p/M_{\oplus} < 30$).

\begin{table*}
\begin{minipage}{17cm}
\begin{center}
\caption{Three zones in the mass-semimajor axis diagram}
\label{table1}
\begin{tabular}{lccc} 
\hline 
Definition$^1$           &  Mass range ($M_{\oplus}$)   &  Semimajor axis range (AU)  & HP13$^2$          \\ \hline 
                                 &                                                 &  0.01 $< r_p <$ 0.1                & Zone 1    \\ [-1ex]
\raisebox{1.5ex}{Hot Jupiters} & \raisebox{1.5ex}{30 $< M_p<$ $10^4$} & 0.1 $< r_p <$ 0.5 & Zone 2 \\[1ex] \hline
Exo-Jupiters              &  30 $< M_p<$ $10^4$            &  0.5   $< r_p <$ 10                 & Zone 3            \\[1ex] \hline
Low-mass planets    &  1 $< M_p<$ 30                      &  0.01   $< r_p <$ 0.5              & Zone 5           \\
\hline 
\end{tabular}

$^1$ the same definitions as HP14.

$^2$ the definitions adopted in HP13.
\end{center}
\end{minipage}
\end{table*}

As shown below, we find that the total PFFs for all the three populations are correlated positively with metallicity. 
This arises from the nature of core accretion picture.
The PFFs of the low-mass planets, which are formed as failed cores of gas giants and/or mini-gas giants in our model, 
correspond to those of the total at up to [Fe/H]$\simeq -1$ above which the population of the Jovian planets becomes non-negligible. 
Our results show that lower metallicity disks tend to create the hot Jupiters more easily than the exo-Jupiters, 
which originates from the combined effects of planet formation with planetary migration. 
We also demonstrate that the exo-Jupiters become dominant over the hot Jupiter population at [Fe/H]$\ga -0.7$. 
We finally examine $\braket{R_{rapid}}$ for the hot Jupiters as well as the low-mass planets, 
which essentially traces the difference in the efficiency of gas accretion between them,  
and find that the critical metallicity for gas giant formation is [Fe/H]$\simeq -1.2$. 
Our results therefore imply that the behavior of the PFFs for different planetary populations and the resultant characteristic metallicities 
may link deeply to important physical processes involved with planet formation. 
Thus,  exoplanet observations around metal-poor stars may contain an invaluable potential to examine these processes.

The plan of this paper is as follows. 
In Section \ref{model}, we briefly describe the semi-analytical model we have employed. 
In Section \ref{resu}, we present our results and derive the critical metallicity above which gas giant formation can take place. 
In Section \ref{disc}, we discuss potential issues that may arise when planet formation around metal-poor stars is considered. 
We also examine how valid our results are by comparing the exoplanet observations that are currently available.
We finally present our conclusions in Section \ref{conc}.
 
\section{Semi-analytical models} \label{model}

We briefly summarize the model adopted in this paper. 
We heavily rely on a formulation developed in a series of papers (HP11, HP12, HP13, HP14), and refer the readers to these papers for details.
Table \ref{table2} summarizes important quantities involved with this work.
Also, all the parameters adopted in this paper are exactly the same as HP14, except for the value of $\eta_{dtg}$ or [Fe/H].

\begin{table*}
\begin{minipage}{17cm}
\begin{center}
\caption{Summary of key quantities in the model}
\label{table2}
\begin{tabular}{ccc}
\hline
Symbol               &   Meaning                                                                                                                                    & Value                                  \\ \hline \hline
                           &   Disk parameters                                                                                                                        &                                             \\  \hline
$f_{dtg}$            &    Dust-to-gas ratio ($=f_{dtg, \odot} \eta_{dtg}$)                                                                    &                                               \\
$f_{dtg, \odot}$  &    Dust-to-gas ratio at the solar metallicity                                                                               & $\simeq 1.8 \times10^{-2}$   \\
$\tau_{dep}$      &    Disk depletion timescale ($=\tau_{dep,0} \eta_{dep} $)                                                                &                                              \\
$\tau_{dep,0}$   &    Typical depletion timescale for the fiducial case                                                                    &   $10^6$ yr                            \\ \hline
                          &    Planetary growth                                                                                                                         &                                               \\ \hline
$M_{c,crit}$       &    Critical mass of cores to initiate gas accretion                                                                       &                                               \\
                          & $=M_{c,crit0} (\dot{M}_c/10^{-6} M_{\oplus} \mbox{yr}^{-1})^{1/4}$                                             &                                               \\
$\dot{M}_c$       &  Accretion rate of planetesimals onto planetary cores                                                                  &                                               \\                                
$M_{c,crit0}$     & a parameter for $M_{c,crit}$                                                                                                          &   $5M_{\oplus}$                     \\
                          & $=10M_{\oplus}f_{c,crit}=10M_{\oplus}(\kappa / 1 \mbox{ cm}^2 \mbox{g}^{-1})^{0.2-0.3}$         &                                               \\
$f_{c,crit}$         & A factor for $M_{c,crit0}$                                                                                                               &  0.5                                        \\
$\kappa$           &  Grain opacities of the envelopes surrounding planetary cores                                                    &  $\simeq 0.1 \mbox{ cm}^2 \mbox{g}^{-1}$ \\
$\tau_{KH}$       &  Kelvin-Helmholtz timescale ($= 10^c \mbox{yr} (M_p/M_{\oplus})^{-d}$)                             &                                                \\   
$(c,d)$               & A set of parameters for $\tau_{KH}$                                                                                             & (9,3)                                        \\
\hline
\end{tabular}
\end{center}
\end{minipage}
\end{table*}

\subsection{Basic model} \label{model1}

The model is designed to compute planet formation and migration in viscously evolving disks with photoevaporation of gas (HP12). 
In order to visualize a series of these processes, 
a set of theoretical evolutionary tracks of growing planets are constructed in the mass-semimajor axis diagram (see Figure \ref{fig1}).
Since the full description of tracks can be found elsewhere (e.g., HP12 or HP14), we focus on the key ingredients in our model.

The essence of the model is to incorporate planet traps that are considered 
as one of the promising solutions to the well-known problem of rapid type I migration for planetary cores \citep[HP11]{mmcf06,mpt07,il08v,hp10,lpm10}. 
Planetary migration arises from resonant, tidal interactions between protoplanets and the surrounding gaseous disks 
out of which the protoplanets are born \citep[e.g.,][]{kn12}.
It is well recognized that type I migration that is effective for cores of gas giants and low-mass planets is very rapid ($\sim 10^5$yr), 
and its direction depends strongly on the disk properties such as the surface density and the temperature of disks. 
As a result, it is important to consider the disk properties in detail to examine the effects of type I migration.
Note that type I migration is distinguished from type II migration that occurs when protoplanets are massive enough to open up a gap in their disks.

The presence of planet traps in protoplanetary disks is a natural consequence of disk structures 
that deviate locally from simple power-law approximation (HP11). 
Due to the high sensitivity of rapid type I migration to disk structures \citep[e.g.,][]{pbck09}, 
such local variations can act as barriers for the migration.
In addition, the location of the traps varys following disk evolution, since the disk properties evolve with time.
As shown analytically by HP11 and numerically by \citet{mpt09}, 
planet traps move inwards on a timescale similar to the disk lifetime ($\sim 10^{6-7}$ years). 
This slow, inward movement of traps is of fundamental importance 
both to prevent planetary cores from plunging into the central stars within the disk lifetime and to form gas giants at $r \ge 1$ AU efficiently (HP12).
Planet traps are effective until trapped protoplanets become massive enough to switch to type II migration. 
We consider three disk properties (dead zones, ice lines, and heat transitions) and associated traps (HP11).

The significance of planet traps can be recognized well when evolutionary tracks of planets forming at planet traps are constructed.
As demonstrated by HP12, the combination of the core accretion picture with planet traps leads to planetary population 
that is consistent with the radial velocity observations of exoplanets 
in a sense that the end points of tracks line up with the observed planets.

We now discuss the core accretion scenario that is adopted as the formulation of planetary growth in the model. 
This scenario consists of two distinct processes: 
the formation of cores, followed by gas accretion onto the cores \citep[HP12, HP14, see also][]{il04i}.
For the core formation, the timescale of the oligarchic growth is used for computing the mass evolution of planetary cores \citep{ki02}. 
Since the timescale decreases with increasing the dust density or [Fe/H] in disks (see Equation (\ref{z_disk})), 
the efficiency of planet formation, especially massive planets, becomes an increasing function of [Fe/H]. 
The feature of the core accretion picture is already shown by the previous studies \citep[HP14]{il04ii,mab12},
and is regarded as one of the clear evidence that the core accretion picture is preferred for understanding the exoplanet observations.

For the gas accretion, we adopt the conservative approach 
that is used in one of the popular population synthesis calculations \citep{il04i,il08,iln13}. 
Practically, gas accretion starts after the mass of planetary cores exceeds the critical value ($M_{c,crit}$) 
beyond which hydrostatic envelopes around the cores cannot be maintained \citep{m80,bp86,ine00,hi10}. 
The mass increase via the gas accretion is then determined by the Kelvin-Helmholtz timescale ($\tau_{KH}$) 
that characterizes the gravitational contraction of the envelopes \citep[e.g.,][]{ine00,il04i}.
The actual formulae for $M_{c,crit}$ and $\tau_{KH}$ are derived from detailed numerical simulations \citep{ine00}.
Note that these two formulae contain 3 fundamental parameters (see Table \ref{table2}): 
$M_{c,crit0}$ is the key parameter to regulate the value of $M_{c,crit}$ 
whereas $\tau_{KH}$ is determined entirely by a set of two parameters, $c$, and $d$.
As shown by HP14, these three parameters can be constrained by the exoplanet observations 
in which the number of exoplanets that are detected by the radial velocity methods is examined as a function of stellar metallicity; 
the best set of them is $M_{c,crit0}=5M_{\oplus}$, $c=9$, and $d=3$.
We adopt these three values for the following calculations.
 
Finally, the metallicity ([Fe/H]) is treated as a parameter in our model and is handled by the following equation 
\citep[][HP14, see also Tables \ref{table2} and \ref{table3} for definition of quantities]{il04ii}:
\begin{equation}
 \label{z_disk}
\mbox{[Fe/H]}  \simeq \mbox{ [m/H]} \equiv \log_{10} \frac{f_{dtg}}{f_{dtg,\odot}} = \log_{10}(\eta_{dtg}),
\end{equation}
where m represents a mixture of metals.
Note that $\eta_{dtg}$ regulates the surface density of dust in disks
that is related to the surface density of gas ($\Sigma_g$) as $\Sigma_d \propto \eta_{dtg} \Sigma_g$.
As pointed out above, the variation of $\eta_{dtg}$ thus affects the efficiency of forming cores of gas giants \citep[][HP14]{il04ii,mab12}.
 
\subsection{The condition for rapid gas accretion} \label{model2}

\begin{figure}
\begin{center}
\includegraphics[width=9cm]{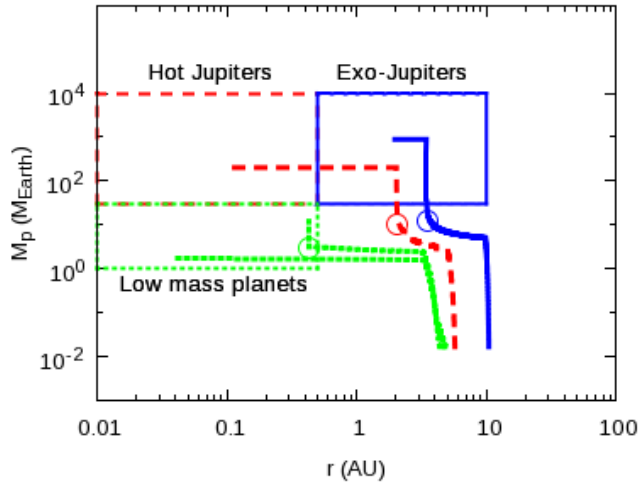}
\caption{Examples of evolutionary tracks of planets growing at planet traps in the mass-semimajor axis diagram.
The diagram is partitioned into three zones where most observed exoplanets distribute (see Table \ref{table1}).  
Tracks evolve with time from a low value of $M_p$ to a large value of $M_p$ (equivalently from a large value of $r$ to a small value of $r$).
The points at which rapid gas accretion starts are denoted by the circles (see Equation (\ref{con_rapid}) for the condition). 
Tracks for the hot Jupiters, the exo-Jupiters, and the low-mass planets are denoted by the dashed, the solid, and the dotted line, respectively. 
Note that one of tracks ending up in the zone of the low-mass planets does not undergo the rapid gas accretion phase 
(so that there is no circle on the track).}
\label{fig1}
\end{center}
\end{figure}

In this paper, the timescale of gas accretion ($\tau_{KH}$) plays an important role.
This is because the fate of growing planets, either the Jovian planets or the low-mass planets, is determined by the efficiency of gas accretion. 
As an example, when the formation of a core completes in the early stage of disk evolution and the core mass is large enough, 
the subsequent gas accretion proceeds efficiently and the core can grow to a gas giant within in the disk lifetime. 
For instance, $\tau_{KH}< 10^6$ yr for $M_p > 10 M_{\oplus}$, which is comparable to or shorter than the typical disk lifetime ($\simeq 10^6-10^7$ yr).
The opposite situation, on the contrary, can be attained if a core needs a long time to initiate gas accretion and its mass is small. 
When $M_p < 1 M_{\oplus}$, $\tau_{KH} > 10^9$ yr, which is beyond the disk lifetime.
This argument therefore indicates that examination of $\tau_{KH}$ can allow differentiation of the Jovian planets from the low-mass planets 
that are formed as failed cores of gas giants and/or mini-gas giants. 
Note that some studies including this work show that a large fraction of low-mass planets in tight orbits such as super-Earths can be generated 
via the same mechanism as forming gas giants, but with the different efficiency of gas accretion \citep[HP12, HP13, HP14, see also][]{rbl11}. 

Here, we focus on the characteristic orbital radius $R_{rapid}$ rather than directly examining $\tau_{KH}$, 
where $R_{rapid}$ is defined as the orbital radius within which the gas accretion becomes efficient enough, 
so that protoplanets can grow up to gas giants. 
This is because examination of $R_{rapid}$ for different planetary populations can be readily achieved 
in computing evolutionary tracks in the mass-semimajor axis diagram (see Figure \ref{fig1}).
Practically, we assume that evolutionary tracks arrive at $R_{rapid}$ if the following condition is satisfied:
\begin{equation}
 \label{con_rapid}
 \tau_{KH}/\tau_{rapid} < 1, 
\end{equation}
where $\tau_{rapid}=10^5$ yr.
We confirmed that the specific choice of $\tau_{rapid}$ does not affect our conclusions 
if $\tau_{rapid}$ is sufficiently shorter than the typical disk lifetime ($\simeq 10^6-10^7$ yr).

Figure \ref{fig1}, as an example, shows how $R_{rapid}$ is estimated along the tracks (see the circles). 
Note that one of the tracks that end up in the zone of low-mass planets does not experience the rapid gas accretion 
and hence no circle on the track (see one of the dotted lines). 

It is important that these examples show that gas giants end up with a larger value of $R_{rapid}$ than low mass planets. 
Such a difference in $R_{rapid}$ for different planetary populations can be understood as follows.
In our model, planet formation is intimately coupled with inward planetary migration, 
either via the inward movement of planet traps or via the inward type II migration.
As a result, the orbital radius of planets shrinks with time, as they grow. 
Since it is preferred to undergo the rapid gas accretion in earlier stage of disk evolution in order to form gas giants efficiently, 
such planets tend to have a larger value of $R_{rapid}$ than planets that eventually fill out the regime of the low-mass planets. 
Note that, as pointed out above, the low-mass planets are formed as failed cores of gas giants and/or mini-gas giants in our model. 
This in turn indicates that planets ending up in the zone of the low-mass planets do not necessarily have $R_{rapid}=0$. 
For the case that $R_{rapid} >0$, the low-mass planets undergo rapid gas accretion, but predominantly in the final disk evolution stage, 
where the disk dissipation proceeds very rapidly, and hence the cores cannot build massive envelopes within the disk lifetime. 
The finally formed planets for this case therefore are regarded as mini-gas giants (see the circle on one of the dotted lines in Figure \ref{fig1}).
For the case $R_{rapid} =0$, on the contrary, the rapid gas accretion never occurs.
As a result, the low-mass planets are considered as failed cores of gas giants with low-mass or negligible envelopes 
(see one of the dotted lines without a circle in Figure \ref{fig1}).

Thus, it is expected that the statistical analysis of $R_{rapid}$ for different planetary populations can provide a useful tool for distinguishing these populations, 
and eventually allow one to quantify the critical metallicity for gas giant formation (see below).
 
\subsection{Statistical approach} \label{model3}

The quantitative comparison with observations is crucial for deriving invaluable constraints on theories of planet formation. 
For this purpose, a new statistical approach was developed in HP13 (see also HP14).
The input parameters for the statistical analysis are summarized in Table \ref{table3}.

\begin{table*}
\begin{minipage}{17cm}
\begin{center}
\caption{Input parameters for the statistical analysis$^1$}
\label{table3}
\begin{tabular}{ccc}
\hline
Symbol             &  Meaning                                                                                                                                     & Values           \\ \hline
                         &  Stellar parameters                                                                                                                      &                                       \\ \hline
$M_*$               &  Stellar mass                                                                                                                                & 1 $M_{\odot}$               \\
$R_*$               &  Stellar radius                                                                                                                                &  1 $R_{\odot}$              \\
$T_*$               &  Stellar effective temperature                                                                                                        & 5780 K                           \\ \hline
                         &  Disk mass parameters                                                                                                                 &                                       \\ \hline
$\eta_{acc}$     &  A dimensionless factor for varying the disk accretion rate ($\dot{M}$)                                         & $0.1 \leq \eta_{acc} \leq 10$  \\
$w_{mass}(\eta_{acc})$ & Weight function for $\eta_{acc}$ modeled by the Gaussian function                                &                                    \\
$\eta_{dtg}$      &   A parameter for varying the metallicity ([Fe/H], see Equation (\ref{z_disk}))                              & $0.01 \la \eta_{dtg} \la 0.25$  \\ \hline
                         &  Disk lifetime parameters                                                                                                               &                                      \\ \hline   
$\eta_{dep}$       &   A dimensionless factor for varying $\tau_{dep}$ (see Table \ref{table2})                                 &  $0.1 \leq \eta_{dep} \leq 10$      \\                                                                                                                                              
$w_{lifetime}(\eta_{dep})$ & Weight function for $\eta_{dep}$ modeled by the Gaussian function                           &                                    \\
\hline
\end{tabular}

$^1$ These quantities are derived from the observations of protoplanetary disks.
\end{center}
\end{minipage}
\end{table*}

This approach starts off from dividing the mass-semimajor axis diagram into zones (see Figure \ref{fig1}, see also Table \ref{table1}).
Such partition is motivated well by the exoplanet observations 
which show that most currently observed exoplanets distribute well in these zones \citep[HP13]{cl13}.

Then, a large number ($N_{int}=300$) of tracks are computed, by focusing on two important disk parameters: 
the disk accretion rate ($\eta_{acc}$) and disk lifetime ($\eta_{dep}$) parameters (see Table \ref{table3}). 
The former one essentially scales the disk mass 
whereas the latter involves the efficiency of photoevaporation that dissipates gas disks totally at the final stage of disk evolution.
In addition to the stellar parameters, population synthesis calculations confirm 
that a set of these three parameters are crucial for regulating planet formation in protoplanetary disks \citep[e.g.,][]{il04i,mab09}.

Counting the number of final points of tracks that end up in a certain zone, Zone i, 
planet formation frequencies (PFFs) are evaluated as (HP13):
\begin{eqnarray}
 \label{pfr}
 \mbox{PFFs(Zone i)}   & \equiv  &  \\  \nonumber   
           \sum_{\eta_{acc}} \sum_{\eta_{dep}} & w_{mass}(\eta_{acc}) & w_{lifetime}(\eta_{dep}) \\  \nonumber        
          & \times &   \frac{N\mbox{(Zone i, } \eta_{acc}, \eta_{dep})}{N_{int}},  \nonumber                           
\end{eqnarray}
where $N\mbox{(Zone i, } \eta_{acc}, \eta_{dep})$ is the total number of tracks that eventually distribute in Zone i after gas disks dissipate,
$N_{int}=300$ is the total number of tracks considered in single calculations, 
and $w_{mass}$ and $w_{lifetime}$ are both weight functions for $\eta_{acc}$ and $\eta_{dep}$, respectively.

The above quantity can be regarded as the PFFs for the following reasons. 
First, the ratio, $N\mbox{(Zone i, } \eta_{acc}, \eta_{dep})/N_{int}$ can be considered 
as the "efficiency" of forming planets in Zone i for a specific set of $\eta_{acc}$ and $\eta_{dep}$. 
Second, the ratio is then integrated over a wide range of both $\eta_{acc}$ and $\eta_{dep}$ with the weight functions, $w_{mass}$ and $w_{lifetime}$.
These two weight functions are represented by the Gaussian function and are formulated by the observations of protoplanetary disks (HP13). 
Specifically, $w_{mass}$ and $w_{lifetime}$ are essentially analytical modeling of 
the observed distribution of the disk accretion rate and the disk lifetime, respectively.
Consequently, the PFFs can be compared with the observations directly without the standard population synthesis calculations being performed.

One of the strong advantages in this approach is to allow one to examine theories of planet formation quantitatively.
As discussed above, HP14 have recently applied this approach to the observational planet-metallicity relation, 
and derived the best set of three fundamental parameters ($M_{c,crit0}$, $c$, and $d$) in the model of planet formation (see Tabel \ref{table2}).
One of the major findings of HP14 is that the canonical value of $M_{c,crit}(\simeq 10M_{\oplus})$ that is widely adopted in the literature 
is likely to be too large to reproduce the observations.

As already discussed in Section \ref{model2}, it is important to estimate the value of $R_{rapid}$ within which the gas accretion becomes efficient.
We calculate $\braket{R_{rapid}(\mbox{Zone i})}$ that is a statistically averaged value of $R_{rapid}$ for tracks ending up at Zone i, 
which is given as:
\begin{eqnarray}
\label{R_runaway}
\braket{R_{rapid}(\mbox{Zone i})} & \equiv & \\  \nonumber   
  \sum_{\eta_{acc}} \sum_{\eta_{dep}} & w_{mass}(\eta_{acc}) &  w_{lifetime}(\eta_{dep}) \\  \nonumber    
                                                            & \times &   \braket{R_{rapid}(\mbox{Zone i, } \eta_{acc}, \eta_{dep})},
\end{eqnarray}
where $\braket{R_{rapid}(\mbox{Zone i, } \eta_{acc}, \eta_{dep})}$ is the mean value of $R_{rapid}$ for tracks 
that eventually freeze in Zone i after the computation is complete, for a specific value of $\eta_{acc}$ and $\eta_{dep}$.

\section{Results} \label{resu}

We present our results in Figures \ref{fig2} and \ref{fig3}. 
We show the resultant PFFs for three different planetary populations and their total, as a function of metallicity, 
and discuss what is the critical metallicity for forming Jovian planets.
Based on our preliminary results, 
we confirmed that it is sufficient to consider a certain range of metallicities ($-2 \le$[Fe/H]$\le -0.6$).

\subsection{The PFFs} \label{resu1}

We first discuss the results of the PFFs for the total as well as the low mass planets, 
and then examine those for both the hot and exo-Jupiters.

\begin{figure}
\begin{center}
\includegraphics[width=9cm]{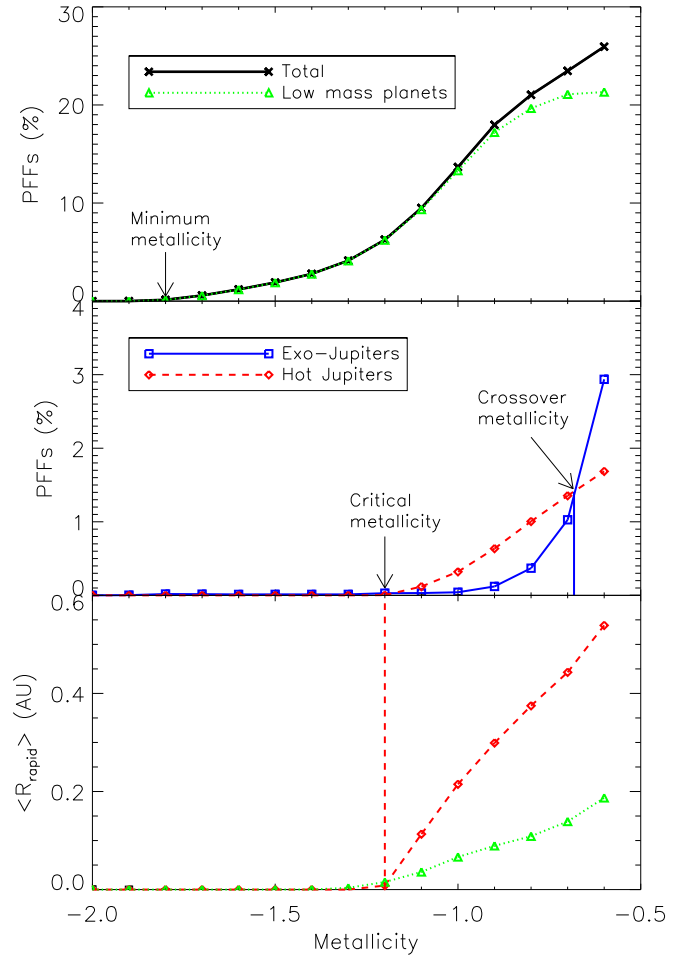}
\caption{The resultant PFFs and mean orbital radius ($\braket{R_{rapid}}$) at which gas accretion becomes rapid ($ \la 10^5$ yr), as a function of metallicity.
The top panel shows the PFFs for the total and the low-mass planets denoted by the thick and dotted lines, respectively 
whereas the middle panel is for the PFFs for the hot and exo-Jupiters denoted by the dashed and solid lines, respectively.
The bottom panel shows the value of $\braket{R_{rapid}}$ for the hot Jupiters as well as the low-mass planets.
The critical metallicity for forming gas giants, that is estimated by the hot Jupiters, 
is denoted by the vertical dashed line (at [Fe/H]$\simeq -1.2$) on the bottom panel. 
The crossover metallicity above which the exo-Jupiters dominate over the hot Jupiters 
is by the vertical solid line (at [Fe/H]$\simeq -0.7$) on the middle panel.
}
\label{fig2}
\end{center}
\end{figure}

Figure \ref{fig2} (Top) shows that the total PFFs steadily increase with metallicity (see the thick line). 
For the low-mass planets (see the dotted line), the PFFs behave similarly to those of the total until the metallicity goes up to [Fe/H]$\simeq -1$. 
Beyond such metallicities, the rapid increase in the PFFs ceases and their value becomes almost constant toward [Fe/H]$\simeq -0.6$.

The behavior of the total PFFs can be understood as a direct reflection of the core accretion scenario. 
As discussed above, the efficiency of planet formation in this picture is correlated positively with the amount of the dust in disks (see equation (\ref{z_disk})). 
Thus, it is evident in our model that more planets are formed for higher metallicity disks.

The deviation of the low-mass planets from the total PFFs originates from the rapid increase in the PFFs for the Jovian planets.
As shown in the middle panel, both the Jovian planets achieve higher values of PFFs at [Fe/H]$\ga -0.7$. 
Consequently, the contribution of gas giants becomes non-negligible around such metallicities. 
In addition, we recall that low-mass planets are formed via the same mechanism as gas giants in our model. 
Specifically, they are regarded as failed cores of gas giants and/or mini-gas giants.
Our results therefore imply that 
the dominant end product starts switching from low-mass planets to massive ones at [Fe/H]$\ga-0.7$ (see the following discussion).

Figure \ref{fig2} (Middle) shows the resultant PFFs for both the hot and exo-Jupiters (see the dashed and solid lines, respectively).
We find that the PFFs for the hot Jupiters are higher than those for the exo-Jupiters at lower metallicity ([Fe/H]$\la -0.7$). 
The results also show that the PFFs for the exo-Jupiters increase with metallicity more rapidly than those for the hot Jupiters.
The crossover occurs at [Fe/H]$\simeq-0.7$ (see the vertical sold line).

We discuss a physical reason of why the formation of the hot Jupiters proceeds more efficiently in lower metallicity disks than that of the exo-Jupiters.
This can be understood as follows.
First, the lower metallicity environments result in the slower rate of planetary growth. 
This is obvious from the core accretion picture \citep[e.g.,][see also HP14]{ki02}.
Second, the effects of planet traps then become more prominent.
As described briefly in Section \ref{model1} and substantially in HP12, 
planet traps can play a crucial role in capturing and moving planetary cores inward. 
Such trapped cores drop-out from their host traps when they obtain the gap-opening mass that is an increasing function of the distance from the central stars.
As a result, when planetary cores take a long time to become massive enough to achieve the gap-opening mass, 
they experience radial drift by their traps over a long distance.
This ends up with the distribution of finally formed gas giants 
that are more likely to fill out the zone of the hot Jupiters than that of the exo-Jupiters in the lower metallicity environment.

It is interesting that the dominant Jovian population switches from the hot Jupiters to the exo-ones at [Fe/H]$\simeq -0.7$.
This occurs because, once the formation of gas giants becomes efficient enough, 
the zone of exo-Jupiters is a more preferred place for gas giants to end up. 
This trend is already shown by HP13 which demonstrate that the resultant PFFs are in good agreement with the exoplanet observations, 
wherein most observed gas giants at $r \le 10$ AU are densely populated around 1 AU with fewer hot Jupiters (see also HP14).

\subsection{The critical metallicity} \label{resu2}

We discuss various characteristic metallicities for forming planets, utilizing the above results.
Our results show that the minimum metallicity for low-mass planets is [Fe/H]$\simeq -1.8$ (see the top panel of Figure \ref{fig2}).
Nonetheless, we focus the following argument only on gas giant formation, 
because low-mass planet formation can also proceed in planetesimal disks that emerge after gas disks totally dissipate, 
which is not modeled in this study.
As a result, the above estimate would provide only an upper limit for the low-mass planets.

Based on the above discussion, 
the critical metallicity for forming gas giants can be estimated intuitively as [Fe/H]$\simeq -1$ (see the middle panel of Figure \ref{fig2}).
In the following, we specify the value more sharply, examining the value of $\braket{R_{rapid}}$ 
(see Sections \ref{model2} and \ref{model3} for the definition). 

Our results suggest that the critical metallicity for gas giant formation will be derived from the hot Jupiters, rather than the exo-Jupiters. 
The main reason is that the population of the hot Jupiters become dominant over the exo-Jupiters at [Fe/H]$\la -0.7$.  
As already pointed out, this occurs due to the lower growth rate of cores in disks with lower metallicities 
and the resultant enhancement of the inward transport of the cores by their traps. 
We refer to the metallicity above which the PFFs for the exo-Jupiters exceed those for the hot Jupiters, 
as the "crossover metallicity" (see Figure \ref{fig2}). 

We now estimate the critical metallicity for forming gas giants, 
focusing on the behavior of $\braket{R_{rapid}}$ for different planetary populations.  
As discussed in Section \ref{model2}, the Jovian planets tend to have a larger value of $R_{rapid}$ than the low-mass planets.
In addition, we recall that the radial extent of the zones of the hot Jupiters and the low-mass planets corresponds with each other (see Figure \ref{fig1}). 
We can therefore point out that the difference in $\braket{R_{rapid}}$ for these two different planetary populations originates from the gas accretion efficiency.
 
\begin{figure}
\begin{center}
\includegraphics[width=9cm]{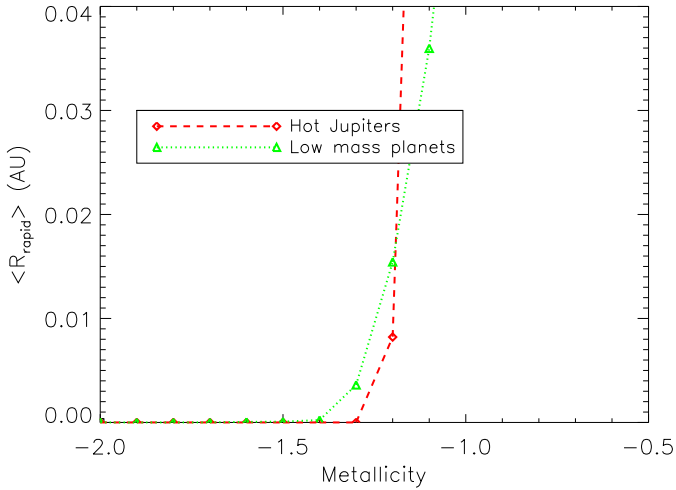}
\caption{A blow-up version of the bottom panel of Figure \ref{fig2}.
The value of $\braket{R_{rapid}}$ for the hot Jupiters exceeds that for the low-mass planets sharply at [Fe/H]$\simeq - 1.2$.}
\label{fig3}
\end{center}
\end{figure}

Figure \ref{fig2} (Bottom) shows the resultant $\braket{R_{rapid}}$ for both the hot Jupiters as well as the low-mass planets 
(see the dashed and dotted lines, respectively, see also Figure \ref{fig3} for a clear demonstration). 
We find that $\braket{R_{rapid}}$ for both populations is zero at up to [Fe/H]$\simeq -1.4$. 
As the metallicity increases, the low-mass planets first achieve a non-zero value of $\braket{R_{rapid}}$ at [Fe/H]$\simeq -1.3$, 
and then the hot Jupiters do at [Fe/H]$\simeq -1.2$. 
Once the value of $\braket{R_{rapid}}$ for the hot Jupiters becomes larger than zero, 
it increases with metallicity more rapidly than that for the low-mass planets.  
The sharp rise occurs at [Fe/H]$\simeq -1.2$.

Our results show that when [Fe/H]$> -1.2$ at which the PFFs for the hot Jupiters are non-negligble, 
$\braket{R_{rapid}}$ for the hot Jupiters obtains a larger value than that for the low-mass planets. 
For the regime of [Fe/H]$< -1.2$, on the contrary, the opposite trend is established, where almost no hot Jupiters are formed (see Figure \ref{fig3}). 
Thus, our results indicate that the critical metallicity for gas giant formation is [Fe/H]$\simeq -1.2$.
This is the major finding of this paper and 
is of quantitative significance in putting a constraint on the core accretion picture for forming massive planets around metal-poor stars 

The resultant behavior of $\braket{R_{rapid}}$ also provides a suggestion for possible members of the low-mass planets.
When [Fe/H]$\leq -1.4$, all the formed planets are failed cores of gas giants, 
so that they are solid cores with low-mass or almost negligible atmospheres. 
As the metallicity goes up, the members of the low-mass planets become the combination of failed cores as well as mini-gas giants 
that contain more masses in their envelopes. 

In summary, we can conclude that various kinds of characteristic metallicities derived here 
are very likely to depend on physical processes in planet formation. 
Specifically, the crossover metallicity can calibrate how planet formation and migration are coupled with each other. 
The examination of $\braket{R_{rapid}}$ for different planetary populations, 
which acts as a crucial agent for quantifying the critical metallicity for gas giant formation, 
can trace the different gas accretion histories between them.
We also note that these metallicities are derived from the model that can well reproduce the current observational data of exoplanets.

\section{Discussion} \label{disc}

As discussed above, we have derived various characteristic metallicities for planet formation (see Figure \ref{fig2}). 
These values may provide a good reference for exoplanet observations around metal-poor stars. 
This is because we have adopted the best choice of the weight functions ($w_{mass}$ and $w_{lifetime}$) 
that can fit well to the observations of protoplanetary disks 
as well as the best set of parameters for planetary growth ($M_{c,crit0}$, $c$, and $d$, see Table \ref{table2}) 
that can reproduce well the statistical properties of currently observed exoplanets ($-0.6 \leq$ [Fe/H] $\leq 0.6$). 
Nonetheless, there may be a number of potential issues that may affect the above values.
We summarize them and discuss how important they are for deriving the critical metallicity.
Also, we discuss how consistent our results are with the current observations of exoplanets around metal-poor stars.
 
\subsection{Metallicity effects on disk evolution}

Disk evolution can be affected by the metallicity. 
This has recently been inferred from the observations 
which show that the disk lifetime in the low metallicity environment is likely to be much shorter than that in the solar metallicity region 
with an $\sim 10^{\mbox{[Fe/H]}}$ dependence \citep{ykt10}. 
This dependence suggests that the disk lifetime at [Fe/H]$=-1$ shortens by a factor of 10, compared with disks with the solar metallicity. 
It is obvious that such short disk lifetimes will make a considerable impact on planet formation.

In our model, photoevaporation, that plays a dominant role in the final stage of disk evolution \citep[e.g.,][]{a11}, 
defines the disk lifetime ($\eta_{dep}$, see Section \ref{model3}).
The above observed disk lifetime-metallicity correlation can be explained marginally by the effects of photoevaporation \citep[e.g.,][]{gh09,ec10}.
It is therefore important to take into account the correlation. 
One of the plausible effects on our results may appear in deriving the critical metallicity for gas giant formation.
As discussed in Section \ref{resu1}, the hot Jupiters are the crucial population for estimating the critical metallicity. 
This is because low metallicity disks prolong the growth timescale of planetary cores, which ends up with efficient inward migration. 
When disk lifetimes become much shorter in the lower metallicity environment, however, 
it can be anticipated that the effects of the inward migration may be diminished. 
This implies that the critical metallicity obtained in Section \ref{resu2} may be underestimated (see Figure \ref{fig2}).
If the disk lifetime-metallicity correlation and the resultant short disk lifetimes would suppress the inward migration more significantly,
then there is a possibility that the exo-Jupiters (but not the hot ones) may become an important population for calibrating the critical metallicity.

We have nonetheless neglected the disk lifetime-metallicity correlation and the resultant non-linear effect in the above calculations.
This is mainly because the statistics of disk observations in the low metallicity environment is still immature.
As a result, it is difficult to make a reliable modeling for $w_{lifetime}$. 
In addition, the primary focus of this work is on a pure effect of the metallicity on the core accretion picture. 

Recently, \citet{jl12} have investigated the critical metallicity for planet formation.
Assuming that the timescale of dust settling provides the most crucial bottleneck for planet formation around extremely metal-poor stars,  
they simply compared the timescale of dust settling with the disk lifetime. 
In their derivation, they take into account the disk lifetime-metallicity correlation discussed above.
They found that the critical metallicity for planet formation is described as [Fe/H]$\simeq -1.5 + \log(r/ 1 \mbox{AU})$.
It can be considered that the condition provides a lower limit for the critical metallicity, 
because dust settling and the subsequent grain growth are prerequisite for the formation of cores of gas giants.
In order to examine the relationship between our results and theirs, we summarize the critical metallicities in Table \ref{table4}. 
Note that we took the radial midpoint of each zone for the orbital radius in the above condition. 
One immediately observes that our results provide larger values than those of \citet{jl12}.
Thus, our results are consistent with theirs 
in the sense that grain growth provides only a necessary condition, but not a sufficient condition for planet formation.

In summary, it is obviously needed to carry out a more intensive observational survey on the disk lifetime around metal poor-stars.
Also, we will undertake a comprehensive theoretical study in which the effect of the the disk lifetime-metallicity correlation is examined in future publication.

\begin{table}
\begin{center}
\caption{Comparison of the critical metallicity for gas giants}
\label{table4}
\begin{tabular}{lccc}
\hline
                          &  Hot Jupiters   & Exo-Jupiters     \\ \hline
This work$^1$   &  -1.2               &  -0.7                    \\
\citet{jl12}$^2$  & -2.6               &  -1.2                   \\

\hline
\end{tabular}

$^1$ For the case of the exo-Jupiters, the crossover metallicity (not the critical metallicity) is labeled (see Section \ref{resu2}).

$^2$ The radial midpoint of each zone is used as the distance from the central star.
\end{center}
\end{table}

\subsection{Metallicity effects on gas accretion}

Planet formation itself is also influenced by the metallicity.
As discussed above, the formation of planetary cores is very sensitive to the metallicity of disks (see also HP14).

One may then wonder how gas accretion onto the cores is affected by the metallicity of surrounding materials that are eventually accreted by the cores. 
If it could be assumed that the opacity in the planetary envelope is directly related to the metallicity of the disk,  
then the disk metallicity would provide an important effect for gas accretion.
This occurs because the critical core mass ($M_{c,crit}$) above which gas accretion initiates depends on the opacity in the envelope (see Table \ref{table2}). 

In our model, $M_{c,crit}$ includes a parameter $M_{c,crit0}$ that is written as \citep{ine00,il04i}
\begin{equation}
\label{mccrit0}
M_{c,crit0}=10M_{\oplus}(\kappa/1 \mbox{ cm}^2 \mbox{g}^{-1})^{0.2-0.3},
\end{equation}
where $\kappa$ is the grain opacity of the envelope surrounding the planetary core.
We have adopted $M_{c,crit0}=5M_{\oplus}$, equivalently $\kappa \simeq 0.1$ cm$^2$ g$^{-1}$ following HP14, 
wherein the value of $M_{c,crit0}$ or $\kappa$ is determined by comparing the currently available exoplanet observations 
that are well confined in a certain range of metallicities ($-0.6 \leq$[Fe/H]$\leq 0.6$).
How can we justify the usage of this fixed value for planet formation in disks around more metal-poor stars ($-2 \leq$[Fe/H]$\leq -0.6$)?

Recently, \citet{mp08} have shown through detailed numerical simulation of planetary envelopes 
that the resultant opacity in the envelope is very unlikely to be sensitive to the accreted materials. 
Namely, the metallicity of disks is very unlikely to determine the efficiency of the gas accretion directly (see equation (\ref{mccrit0})).
Instead, the growth of dust grains and their subsequent settling in the envelope play a more important role in regulating the gas accretion \citep{mbp10}.
If this is the case, we can use the same value of $M_{c,crit0}$ as above for examining the formation of first gas giants. 
In other words, the resultant opacity, $\kappa \simeq 0.1$ cm$^2$ g$^{-1}$, which is about one order of magnitude lower than the canonical value, 
is considered as a consequence of grain growth and subsequent settling in the envelope surrounding planetary cores, 
rather than the reflection of the accreted materials with low metallicities.

Thus, it is very unlikely that gas accretion adopted in our model affects our results significantly.\footnote{
We recall that our model adopts the core accretion scenario, so that the above argument can be applied. 
When gravitational instability scenario, in which planets are formed directly from gravitationally unstable disks like stars, 
would be more important to account for planet formation around metal-poor stars, 
the disk metallicity may play a more significant role \citep[e.g.,][]{b02,cdm06,mb10}.}

\subsection{Relation between $\eta_{dtg}$ and [Fe/H]}

In the above calculations, 
we have assumed a simple relationship between the dust-to-gas ratio ($\eta_{dtg}$) and the metallicity ([Fe/H]) (see equation (\ref{z_disk})). 
On the contrary, there is currently accumulating evidence that $\eta_{dtg}$ does not scale simply with [Fe/H]. 
In fact, both theoretical and observational studies suggest 
that $\eta_{dtg}$ is very likely to drop more rapidly than [Fe/H] at low metallicity \citep[e.g.,][]{lf98,h99,i03,ath13a,rmg14,fbh14}.
Note that this non-linear effect may be valid predominantly in low gas density regimes. 
In dense gas regions such as protoplanetary disks, gas-phase metals may be accreted onto dust grains \citep{cny13}, 
which tends to weaken the effect.
Nonetheless, it is interesting to examine the effect of nonlinear $\eta_{dtg}$-[Fe/H] relation on the critical metallicity of planet formation.

In order to proceed, 
we utilize the results of \citet{ath13a}, wherein dust formation history in a galaxy is investigated, 
focusing on the metal enrichment through galaxy evolution.
Digitizing their Figure 3 in which the dust-to-gas ratio is plotted as a function of the metallicity,  
we can estimate the corresponding metallicity that is computed by their model, based on our results of $\eta_{dtg}$.
Specifically, we first obtain the value of $\eta_{dtg}$ that provides the critical metallicity in our calculations (see equation (\ref{z_disk})), 
and then we, based on the value of $\eta_{dtg}$, read the corresponding metallicity using Figure 3 of \citet{ath13a}.
Note that their results are parameterized by the star formation timescale $\tau_{SF}$. 
This is because the metal enrichment and the subsequent dust formation are involved with 
the formation and evolution of stars via stellar nucleosynthesis and stellar ejecta.

Table \ref{table5} summarizes the corresponding metallicities. 
One observes that the corresponding metallicities become larger for a lower value of $\tau_{SF}$. 
This occurs because shorter timescales of $\tau_{SF}$ end up with more rapid formation of stars. 
As a result, the metal enrichment proceeds more efficiently 
before significant dust growth occurs in the interstellar medium.

This kind of parameter study is interesting in a sense that 
the observations of exoplanets around metal-poor stars may have a potential to put a new constraint on the star formation rate in galaxies.
We will undertake a more complete study about the relation in future publication.

\begin{table}
\begin{center}
\caption{Corresponding metallicities$^1$}
\label{table5}
\begin{tabular}{lcccc}
\hline
                                   &  Minimum                    & Critical                  &  Crossover                     \\ 
                                   &  metallicity                   & metallicity             &   metallicity                     \\ \hline
This work                   &   -1.8                            & -1.2                       &   -0.7                                \\
$\tau_{SF}=0.5$Gyr    &  -0.3                             &  -0.2                      &   -0.1                             \\
$\tau_{SF}=5$Gyr      &  -0.8                             &  -0.7                      &   -0.6                              \\
$\tau_{SF}=50$Gyr    &  -1.3                             &  -1.1                      &   -0.8                             \\

\hline
\end{tabular}

$^1$ In order to examine the effect of nonlinear $\eta_{dtg}$-[Fe/H] relation on the critical metallicity, 
we adopt Figure 3 of \citet{ath13a} in which the theoretically computed dust-to-gas ratio is plotted as a function of metallicity.
Based on our results of $\eta_{dtg}$, the corresponding metallicity is read from their Figure 3.
Since their results are parameterized by the star formation timescale, $\tau_{SF}$,
the corresponding metallicities for three different values of $\tau_{SF}$ are shown.

\end{center}
\end{table}

\subsection{Implications for the exoplanet observations}

Taking into account the above caveats, 
we now discuss implications of our results for exoplanets that are currently observed around metal-poor stars.

As discussed in Section \ref{intro}, the discoveries of exoplanets at [Fe/H]$\simeq -2$ are currently under debate. 
Our results also suggest that it is very difficult to form gas giants at such values of metallicity.

Another example may be an exoplanet observed around a pulser in Messier 4 globular cluster \citep{srh03}. 
The original paper stated that the metallicity of the cluster is about -1.3, which contradicts our prediction. 
Nonetheless, the more recent study infers that the metallicity of the cluster is very likely to be about -1 \citep{mvp08}.
If this is the case, our results are still valid to account for the presence of such a planet.

We also explore exoplanets around metal-poor stars, using the standard archives for exoplanet data; 
the Extrasolar Planets Encyclopaedia \citep[http://exoplanet.eu/]{sdl11} and the Exoplanet Orbit Database \citep[http://www.exoplanets.org]{wfm11}.
We however do not find any example which invalidates our prediction so far.

It is obviously important to perform a more comprehensive observational survey for detecting exoplanets around metal-poor stars, 
and to examine our results.
     
\section{Conclusions} \label{conc}

We have quantitatively investigated various characteristic metallicities for planet formation 
that was motivated by the recent success of exoplanet observations.
To achieve such a goal, we have adopted a formalism developed in a series of earlier papers (HP11, HP12, HP13, HP14). 
The main features of the model are planet traps at which rapid type I migration for planetary cores is halted. 
Three types of planet traps (dead zone, ice line, and heat transition) have been considered. 
For planetary growth, we have relied on the standard core accretion scenario, 
wherein the formation of planetary cores is sensitive to the dust density in disks or the metallicity ([Fe/H]). 
Coupling the scenario with the planet traps in viscously evolving disks with photoevaporation 
has enabled one to compute the evolution of planets in the mass-semimajor axis diagram (see Figure \ref{fig1}). 

We have utilized a set of theoretical evolutionary tracks of planets 
for deriving the planet formation frequency (PFF) as well as the statistically averaged orbital radius 
at which growing protoplanets undergo rapid gas accretion ($\braket{R_{rapid}}$, see Figure \ref{fig1}). 
Specification of $R_{rapid}$ is crucial in this study, because it allows one to estimate the critical metallicity for gas giant formation. 
This was achieved by dividing the mass-semimajor axis diagram into zones as well as counting the end points of tracks for each zone.
Zoning the diagram is in fact inferred from the accumulation of observed exoplanets. 
We have considered three different planetary populations: hot and exo-Jupiters, and low mass planets (see Table \ref{table1}).

We have obtained the PFFs for three types of planets as well as the total, as a function of metallicity (see Figure \ref{fig2}). 
We have shown that the total PFF is a steady increasing function of metallicity, which can be understood straightforwardly by the core accretion scenario.
The PFFs of the low-mass planets follow those of the total until the population of the Jovian planets provide a significant contribution, 
which occurs at [Fe/H]$\ga -1$. 
We have also computed the PFFs for both types of the Jovian planets which show 
that the hot Jupiters can form at lower metallcities than the exo-Jupiters. 
Switching of the dominant population between these two types is a consequence of the intimate coupling of planet formation and migration,
and takes place at [Fe/H]$\simeq -0.7$ that is referred to as the crossover metallicity in this paper.

We have also estimated various characteristic metallicities for planet formation (see Figure \ref{fig2}, see also Table \ref{table5}).
The minimum metallicity for low-mass planets is [Fe/H]$\simeq - 1.8$, although it probably provides only an upper limit (Section \ref{resu2}).
We have quantified the critical metallicity for gas giant formation. 
This was derived from the behavior of $\braket{R_{rapid}}$ 
that essentially links to the different efficiency of gas accretion between the hot Jupiters and the low-mass planets.
We have found that [Fe/H]$\simeq - 1.2$ is the critical metallicity for forming gas giants around metal-poor stars.
This is a most important finding of this study.
We have also pointed out that these values of metallicity depend on the processes of planet formation and migration. 
Our results therefore infer that the observations of exoplanets around metal-poor stars may be used as a probe for calibrating of these processes,   
and hence such observations are very important for examining the current understanding of planet formation in protoplanetary disks.

We have discussed a number of potential issues that may affect our results (Section \ref{disc}).
Specifically, we have examined the effects of metallicity on disk evolution as well as planet formation.
The recent observations suggest that the reduction in metallicity shortens the disk lifetime significantly. 
It is obvious that the effect prevents planet formation. 
Nonetheless, we have neglected it, because more intensive observations are required to make a reliable modeling.
Planet formation is also affected by lowering the metallicity. 
We have discussed the effects on gas accretion. 
It is important that the resultant opacity in planetary envelopes is unlikely to depend strongly on the metallicity of materials accreted by planetary cores.
Thus, the setup adopted in this paper may be sufficient.
 
We have also examined the non-linear effect on the relationship between the dust-to-gas ratio ($\eta_{dtg}$) and the metallicity. 
This was motivated by the recent observational progress on the relationship in the lower metallicity environment, 
wherein $\eta_{dtg}$ is very likely to decrease more rapidly than [Fe/H].
We have calibrated the effect, making use of one of the most recent results in which it is investigated how the relationship evolves with time.
We have found that the corresponding metallicities that are derived from theoretical nonlinear $\eta_{dtg}$-[Fe/H] relation
depend sensitively on the star formation timescale (see Table \ref{table5}). 
This is because the formation and evolution of stars play an important role in enriching the metallicity.
Our results therefore may be useful for bridging the observations of exoplanets around metal-poor stars 
with the estimate of the star formation history through galaxy evolution.

We have finally provided some implications for exoplanets around metal-poor stars that are currently observed so far. 
There is no example which violates our prediction at the moment.
It is however obvious that more observational data are required to examine our results.

In a subsequent paper, we will undertake a more comprehensive study in which the effect of stellar masses will be examined. 
Also, we will consider galaxy evolution simultaneously, which triggers the change of metallicity with time. 
Thus the study will allow a more complete discussion about when the first planets form in galaxies. 


\acknowledgments
The authors thank Ralph Pudritz for stimulating discussion, Ryosuke Asano for providing us with the data of dust enrichment in galaxies, 
and an anonymous referee for useful comments on our manuscript.
Y.H. is supported by EACOA Fellowship that is supported by East Asia Core Observatories Association which consists of 
the Academia Sinica Institute of Astronomy and Astrophysics, the National Astronomical Observatory of Japan, the National Astronomical 
Observatory of China, and the Korea Astronomy and Space Science Institute.  
H.H. is supported by NSC grant: NSC102-2119-M-001-006-MY3.






\bibliographystyle{apj}          

\bibliography{apj-jour,adsbibliography}    

\end{document}